\documentclass[letter]{aa}  
\usepackage{bm}
\usepackage{graphicx}
\usepackage{txfonts}
\usepackage[colorlinks=true,linkcolor=blue,citecolor=blue,urlcolor=blue]{hyperref}
\usepackage[dvipsnames]{xcolor}
\usepackage[normalem]{ulem}
\usepackage{soul}

\newcommand{\mdot}{\ensuremath{\dot{M}}}
\newcommand{\Msun}{\ensuremath{\,M_{\odot}}}	
	
\newcommand{\vinf}{\ensuremath{\varv_{\infty}}}	
\newcommand{\kms}{\ensuremath{\,~\textrm{km\,s}^{-1}}}	
\newcommand{\teff}{\ensuremath{T_{\mathrm{eff}}}}
\newcommand{\Vcl}{\ensuremath{\bar{V}_{\mathrm{cl}}}}
\newcommand{\Mdotcl}{\ensuremath{\dot{M}_{\mathrm{cl}}}}
\newcommand{\Lmecht}{\ensuremath{L_{\textrm{mech, tot}}}}

\usepackage{xcolor}
\usepackage{ifthen}

\usepackage{etoolbox}
\usepackage{lineno}

\begin{document} 
\nolinenumbers
\authorrunning{
    Larkin et al.}

    \title{Mass loading of outflows from evolving Young Massive Clusters}

   \author{C. J. K. Larkin\inst{\ref{inst:mpik},\ref{inst:ari},\ref{inst:mpia}}
          \and
            C. Hawcroft\inst{\ref{inst:stsci}}
          \and
          J. Mackey\inst{\ref{inst:dias}}
          \and
            R. R. Lefever \inst{\ref{inst:ari}}
          \and
          L. H\"arer\inst{\ref{inst:mpik}}
          \and
          A. A. C. Sander\inst{\ref{inst:ari},\ref{inst:iwr}}
                    }

   \institute{{Max-Planck-Institut f\"{u}r Kernphysik, Saupfercheckweg 1, D-69117 Heidelberg, Germany\label{inst:mpik}}\\
            \email{cormac.larkin@mpi-hd.mpg.de}
         \and 
         {Zentrum für Astronomie der Universität Heidelberg, Astronomisches Rechen-Institut, M{\"o}nchhofstr. 12-14, 69120 Heidelberg, Germany\label{inst:ari}}
         \and 
         {Max-Planck-Institut f\"{u}r Astronomie, K\"{o}nigstuhl 17, D-69117 Heidelberg, Germany\label{inst:mpia}}
         \and 
        {Space Telescope Science Institute, 3700 San Martin Drive, Baltimore, MD 21218, USA\label{inst:stsci}}
         \and
          {Astronomy \& Astrophysics Section, School of Cosmic Physics, Dublin Institute for Advanced Studies, DIAS Dunsink Observatory, Dublin D15 XR2R, Ireland\label{inst:dias}}
         \and
    {Universit\"at Heidelberg, Interdisziplin\"ares Zentrum f\"ur Wissenschaftliches Rechnen, 69120 Heidelberg, Germany\label{inst:iwr}}
                      }

   \date{Received DATE; accepted DATE}

  \abstract{Feedback from Young Massive Clusters (YMCs) is an important driver of galaxy evolution. In the first few Myr, mechanical feedback is dominated by collective effects of the massive stellar winds in the YMC. The mass-loss rates and terminal wind velocities of these stars change by orders of magnitude over pre-SN timescales as the massive stars evolve, and mass-loss rates of Cool Supergiant (CSG) stars in particular are uncertain by a factor $\sim$20 or more.
  In this work we perform a first study of the time evolution of average cluster wind velocity \Vcl\ as a function of stellar metallicity $Z$, assuming single star evolution. We also check the validity of assuming Wolf-Rayet stars dominate the feedback effects of a YMC, as often done when interpreting X-ray and $\gamma$-ray observations, and test how sensitive \Vcl\ is to current uncertainties in mass-loss rates.
  We use \textsc{pySTARBURST99} to calculate integrated properties of YMCs for $Z$ in the range of 0.0004--0.02, representing a range of environments from IZw18 to the Galactic Centre.
We find that \Vcl\ drops off rapidly for sub-LMC $Z$, and we recommend a value of 500-1000\kms\ be used in this regime. We show accounting only for WR stars can overestimate \Vcl\ by 500-2000\kms at $Z \geq Z_\text{LMC}$. We also find that different RSG mass-loss assumptions can change the inferred \Vcl\ by $\sim1000$\kms, highlighting the need for improved observational constraints for RSGs in YMCs.
  }


   \maketitle
\nolinenumbers
\section{Introduction}
Young massive stellar clusters (YMCs) are key drivers of stellar feedback in galaxies, hosting the majority of massive stars which mediate this feedback via strong winds during their lifetimes, and deaths as core collapse supernovae (CCSNe) \citep[e.g.,][]{PortegiesZwart2010}. They are commonly found in energetic environments such as luminous infrared- and starburst galaxies, and with the advent of JWST there is a new wealth of observations for YMCs and their feedback across a range of metallicities $Z$ \citep[e.g.,][]{Linden2024ApJ...974L..27L,Kader2025,Correnti2025arXiv250703420C}.

CCSNe feedback begins after only a few Myr. The start time is uncertain as explodability depends on multiple stellar structure variables \citep[e.g.,][]{Burrows+2024,Maltsev+2025}, which themselves are strongly influenced by $Z$-dependent stellar evolutionary effects. \citet{FICHTNER2024} studied feedback from a large grid of stellar evolution calculations, using the SN explodability and explosion energy prescription of \citet{Schneider2021}, for low-$Z$ stellar populations. They found that SNe begin to dominate the IMF-weighted mechanical luminosity after 5\,Myr in all cases, as more massive stars that have shorter lifetimes than this are not predicted to explode.

Pre-SN feedback depends greatly on the individual stellar mass-loss rates \mdot\ and terminal wind velocities \vinf, both of which vary by orders of magnitude for massive stars during post-main sequence evolution. The O and Wolf-Rayet (WR) stars drive a collective outflow with their hot, powerful winds, which are sensitive to $Z$ as well \citep[e.g.,][]{Sander+2020,Hawcroft2024}. Cool Supergiants (CSGs) have much slower winds but similar or higher \mdot\ \citep[e.g.,][]{Yang2023}. While there is good correspondence between wind prescriptions for OB and WR stars, underpinned by detailed theory, this is not the case for CSGs. The lack of coherent theory for CSG \mdot\, and difficulties in direct measurements lead to large uncertainties \citep[e.g.,][]{Beasor2020,Antoniadis+2025,Bronner+2025}, with different assumptions leading to drastic consequences for both subsequent stellar evolution and resulting feedback \citep{Merritt2025}. Values for \vinf\ are comparatively well-constrained and are only weakly dependent on $Z$ \citep{Hawcroft2024}. 

Pre-SN YMC feedback depends on individual stellar winds collectively driving a cluster wind with average velocity, \Vcl, total mechanical velocity, \Lmecht, and total mass-loss rate, \Mdotcl. \citet{Canto2000} presented an analytic model for \Vcl\ assuming maximally efficient wind-wind interactions and mass-loading from individual stars, based on the model \citet{Chevalier1985} developed for starburst galaxies. \citet{Stevens2003} compared expected cluster wind properties with X-ray observations with a similar approach. Following \citet{Canto2000} and \citet{Stevens2003}, we can express \Vcl\ as

\begin{equation}
    \bar{V}_{\mathrm{cl}} =\sqrt{\frac{\sum_{i=1}^N \dot{M}_i V_i^2}{\sum_{i=1}^N \dot{M}_i}} \equiv \left(\frac{2 L_{\textrm{mech, tot}}}{\dot{M}_{\mathrm{cl}}}\right)^{1/2}\text{.}
\end{equation}
\Vcl\ is often used to approximate the cluster wind velocity field  when modelling X-ray and $\gamma$-ray emission from YMCs and their immediate surroundings \citep[e.g.,][]{Morlino2021,Albacete2023ApJS..269...14A,Haerer2023,Webb2024,Haubner2025}. In these cases, it is often assumed that $L_{\textrm{mech, tot}} \approx L_{\textrm{mech, WR}}$, the mechanical luminosity of the WR population, and thus the value of \Vcl\ assumes only this population, neglecting mass-loading from other stars. Recent local-scale (M)HD simulations of young clusters with individually resolved stars \citep[e.g.][]{Badmaev2022,Vieu2024,Haerer2025} have been focussed on the dynamics on pc-scales and the ability to accelerate particles to (very) high energies, therefore largely omitting the winds of cool, evolved stars. Mass-loading of a hot, fast cluster wind through mixing with the cool, slow wind of a single RSG was studied by \citet{Larkin2025}, corresponding well with observations of the RSGs in Westerlund 1 \citep{Guarcello2024}.

Given the high temperatures and low densities, there is a lack of direct \Vcl\ observations. Only the flow speed in the core is indirectly constrained, as it should not significantly exceed the adiabatic sound speed, inferred from diffuse X-ray observations. There is some tension between these values and \Vcl\ predicted by theory, which tend to be a factor $\sim2$ larger \citep{Stevens2003,Haerer2023,Larkin2025}. More direct measurements of \Vcl\ could be done where CSG and cluster winds interact. IFU spectroscopy of the interaction region could reveal \Vcl\ through balancing with ram-pressure \citep[e.g.,][]{Povich2008,Larkin2025}. Yet, no such observations exist.

No studies to date have examined the variation of cluster wind properties with $Z$, despite strong variations in both \mdot\ and \vinf\ with this parameter, and the importance of YMC feedback in the low-$Z$ early Universe. In this work we use population synthesis to (a) compare \Vcl\ and \Lmecht\ for a range of $Z$, (b) compare estimates of \Vcl\ assuming only WR stars vs.~a complete stellar population, and (c) test how sensitive \Vcl\ is to the choice of stellar wind prescription, with a focus on CSGs. We do not consider SNe in this work as we focus on the earliest stages of YMC lifetimes, before SNe become dominant for YMC feedback. 

Our paper is organised as follows: In Sect. \ref{sec:methods} we describe our framework and stellar wind prescriptions. In Sect. \ref{sec:results} we describe $Z$-dependent trends in $V_{\mathrm{cl}}$ and $L_{\textrm{mech, tot}}$ for YMCs (\ref{sec:res_lmech_Vcl}), differences in $V_{\mathrm{cl}}$ for complete stellar populations compared to only WR stars (Sect. \ref{sec:res_all_WR}) and the influence of different stellar wind prescriptions on $V_{\mathrm{cl}}$ (Sect. \ref{sec:res_Vcl_prescr}). We discuss our findings and present our conclusions in Sect. \ref{sec:discussion}.

\section{Methods}
\label{sec:methods}

\subsection{pySTARBURST99}
In this work we use \textsc{pySTARBURST99}, hereafter \textsc{pySB} \citep{Hawcroft2025}, which is an updated version of \textsc{STARBURST99} \citep{Leitherer1999,Leitherer2014}, a population synthesis code intended for calculating the integrated properties of a starburst galaxy. This code calculates values such as \Lmecht\ and \Mdotcl\ for a single coeval stellar population of a given $Z$. 

For all populations modelled here, we take $M_{\mathrm{tot}}=1\times10^{5}$ \Msun\ and a maximum initial stellar mass ($M_{\mathrm{init}}$) of 120 \Msun. We calculate models for all $Z$ values in \textsc{pySB} except zero. These represent $Z$ typical of the Galactic Centre ($Z_\text{MWC}$=0.02), the Galaxy ($Z_\text{MW}$=0.014) the Large and Small Magellanic Clouds ($Z_\text{L/SMC}$=0.006, 0.002) and IZw18 ($Z_\text{IZw18}$=0.0004). For each $Z$, we adopt the non-rotating evolutionary tracks. Using model parameters from these tracks, at each timestep we assign each star to a stellar class, following those used in \textsc{pySB}:
\begin{enumerate}
    \item LBV: $3.75 < \log_{10} T_{\mathrm{eff}} < 4.4$ \& $\log_{10} \dot{M} > -3.5$
    \item CSG: $\log_{10} T_{\mathrm{eff}} < 3.9$ and not an LBV
    \item WR: $\log_{10} T_{\mathrm{eff}} > 4.4$ \& surface H abundance $< 0.4$
    \item OB: Remaining stars with $\log_{10} T_{\mathrm{eff}} > 3.9$
\end{enumerate}

Where we implement new RSG \mdot~ prescriptions, we apply them to CSGs where $\log_{10}T_{\mathrm{eff}} < 3.7$ and $M_{\mathrm{init}}>8$\Msun~, with remaining CSG \mdot~ from the base prescription. This matches where RSG \mdot~ is used in the evolutionary models. Stars with $M_{\mathrm{init}}$ less than $7\,M_\odot$ do not lose mass on the main sequence in these evolutionary tracks, so these filters cover all stars that are relevant for this work.

\subsection{Stellar Wind Prescriptions}

\textsc{pySB} includes a default wind prescription \citep[which we refer to as ``base'', see][for details]{Hawcroft2025}, as well as options for using \mdot\ rates from \citet{Vink2021} (Vink) or \citet{Bjorklund2021} (Leuven), or the \vinf\ relation of \citet{Hawcroft2024} (XShootU), for the main sequence OB stars. We also add the \mdot\ prescriptions of \citet{SV2020,Sander2023} (Sander) for WRs. Given the large number of possible RSG wind prescriptions \citep[see][and references therein]{Yang2023,Decin2024}, we implement recent examples of a low \citep{Beasor2020,Beasor2023} and a high \citep{Yang2023} \mdot\ rate. We do not include the theoretically motivated prescription of \citet{KEE2021} given the range of possible \mdot\ values for reasonable turbulent velocity choices \citep{Merritt2025}. The effects of the alternative \mdot\ prescriptions on the stellar evolution are not accounted for, and so the positions in the HR diagram may not be consistent at all times.

\section{Results}
\label{sec:results}
\subsection{Metallicity trends for $L_{\text{mech, tot}}$ and $V_{\mathrm{cl}}$}
\label{sec:res_lmech_Vcl}

\begin{figure}
    \centering
    \includegraphics[width=1\linewidth]{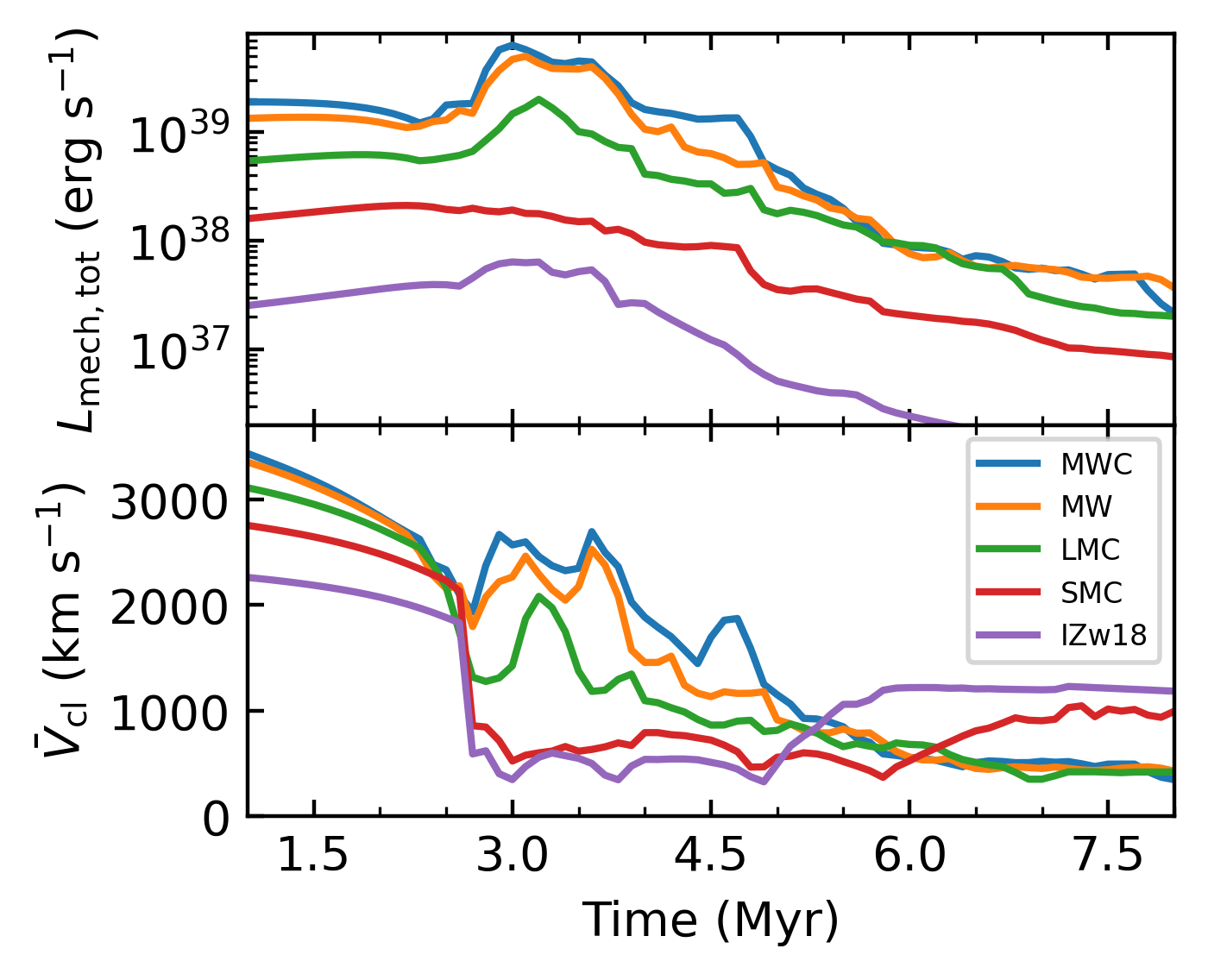}
    \caption{Time evolution of $L_{\textrm{mech, tot}}$ (upper panel) and \Vcl\ (lower panel) for different $Z$.}
    \label{fig:lmech_vcl}
\end{figure}

In Fig.~\ref{fig:lmech_vcl} we show \Lmecht\ and \Vcl\ for different $Z$. The upper panel shows a consistent \Lmecht\ trend. The variation is smooth during the pre-WR phase until $\sim$2.5\,Myr. At this time, \Lmecht\ increases for the MWC, MW, and LMC populations due to the effects of WR stars. These effects are weaker at $Z_\text{SMC}$ and lower, as the stellar evolution tracks (themselves Z-dependent) produce few or no WR stars as defined in \textsc{pySB}. We discuss this effect in App. \ref{sec:OB_WR}. Peak \Lmecht\ occurs at $\sim3$\,Myr in all cases except the SMC, and decreases rapidly as WR stars die off, as shown in \citet{Hawcroft2024}. 

The lower panel shows a qualitatively consistent trend of slowly decreasing \Vcl\ until $\sim$2.5\,Myr. This is where evolved massive stars begin to affect the population, and there is a clear separation from this time forward. For the higher $Z$ populations, \Vcl\ increases due to WR stars, whereas at lower $Z$ \Vcl\ falls off rapidly due to increased mass loading from cool evolved stars, which is not offset by powerful WR winds.  

\subsection{Effects of complete stellar populations}
\label{sec:res_all_WR}

\begin{figure}
    \centering
    \includegraphics[width=1\linewidth]{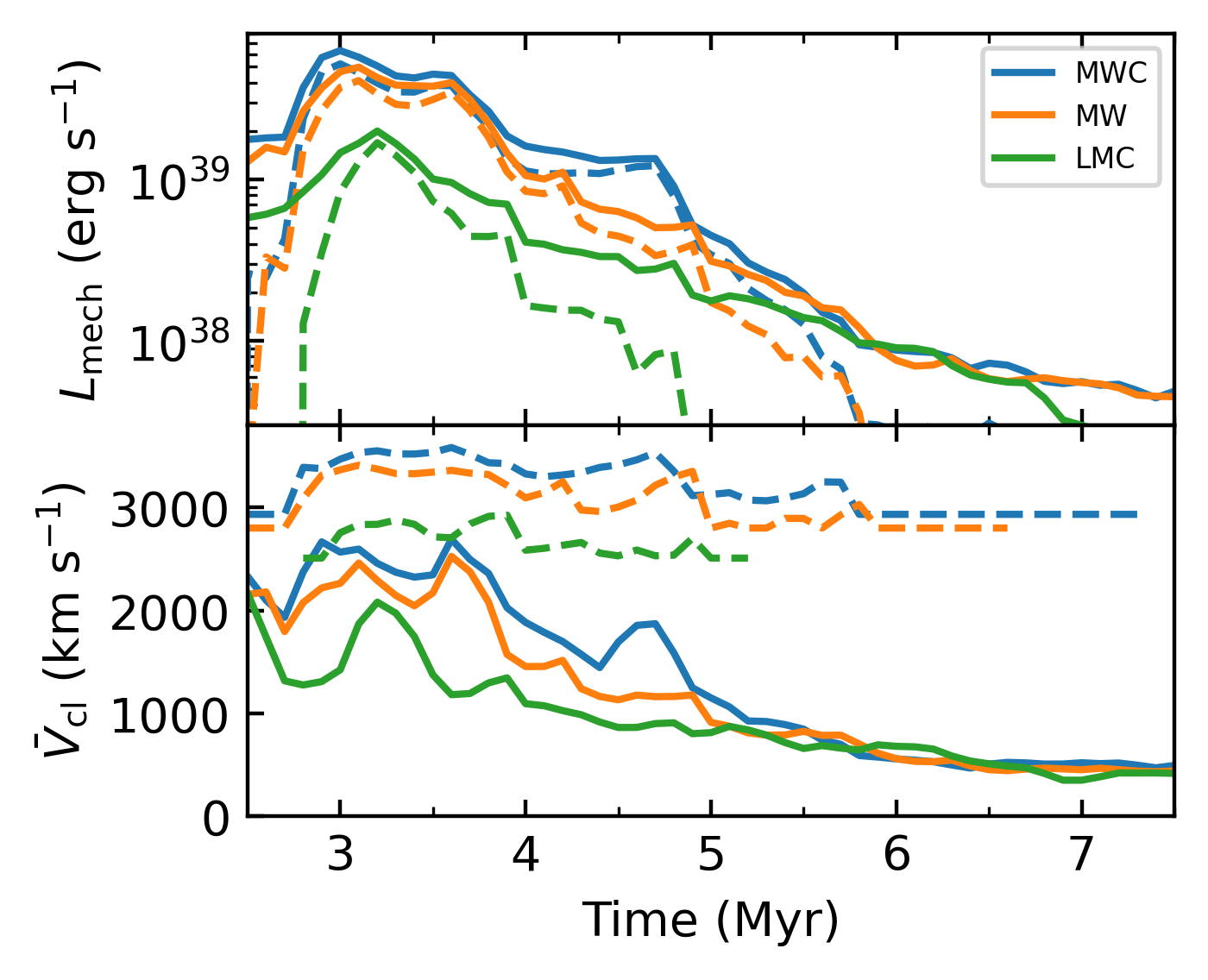}
    \caption{Time evolution of $L_{\mathrm{mech}}$ (upper panel) and \Vcl\ (lower panel) for different $Z$, assuming all stars (solid line) or only WR stars (dashed line).}
    \label{fig:all_vs_WR}
\end{figure}

In Fig.~\ref{fig:all_vs_WR} we show \Lmecht\ and \Vcl\ for $Z_\text{LMC}$ and higher, where single WR stars are relevant. The upper panel validates the canonical assumption that $L_{\textrm{mech, tot}} \approx L_{\textrm{mech, WR}}$. This is generally true for the time when WR stars are active, with a gradual divergence at later times for the LMC population in particular. When active, WR stars account for $\sim$$80\%$ of $L_{\textrm{mech, tot}}$, with the OB stars contributing almost the entirety of the remainder. 

\begin{figure}
    \centering
    \includegraphics[width=1\linewidth]{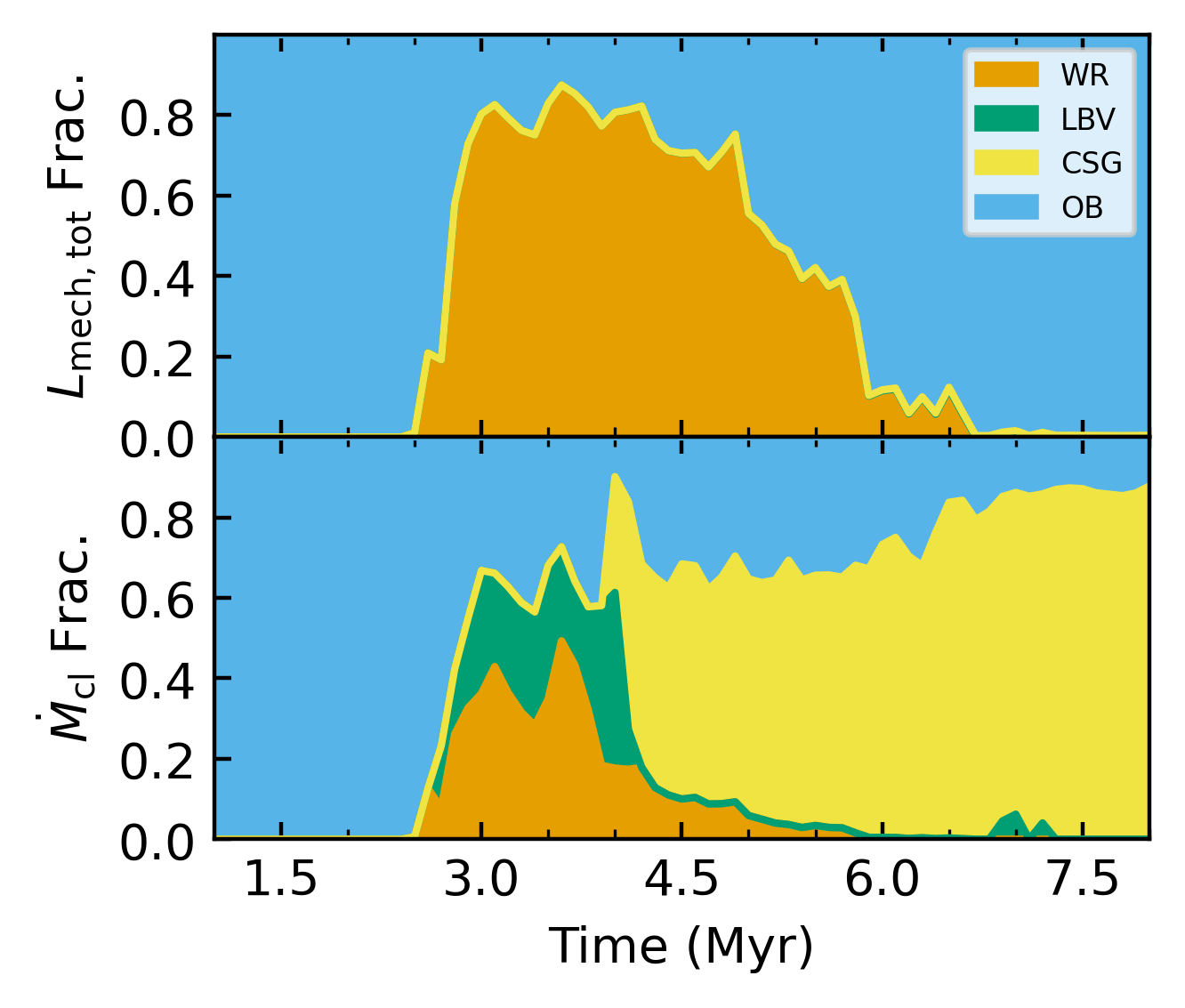}
    \caption{Time evolution of relative contribution of each stellar class to total \Lmecht\ (upper panel) and \Mdotcl\ (lower panel) for $Z_\text{MW}$.}
    \label{fig:lmech_mdot_spt}
\end{figure}

In the lower panel, we see the effects of mass-loading from non-WR stars in these populations. Assuming only WR stars, $V_{\mathrm{cl}}\sim2500-3500$\kms\ over the time period they are present. Including all stars, there is an initial peak from WR winds of $V_{\mathrm{cl}}\sim2500-3000$\kms\ which rapidly falls off. From $\sim$4\,Myr WR stars begin to die off and cool evolved stars become active. This leads to \Lmecht\ decreasing and \Mdotcl\ increasing simultaneously, and thus \Vcl\ falls off. Even for early times, the WR-only \Vcl\ exceeds that of the full population by $>500$\kms. By $\sim4-5$\,Myr, the age typically assumed for the prototypical Galactic YMC Westerlund 1, the difference increases to $\sim$1500\kms. This is because WR stars do not dominate \Mdotcl\ as they do \Lmecht\ for the time they are active -- WR \mdot\ is typically 30\% of total \mdot\ and never exceeds 60\% of it. We show how different stellar classes contribute to \Lmecht\ and \Mdotcl\ over time for Galactic $Z$ in Fig. \ref{fig:lmech_mdot_spt}.

\subsection{Cluster wind sensitivity to prescription}
\label{sec:res_Vcl_prescr}
We now study effects of different stellar wind prescriptions, all at $Z_\text{MW}$. For the OB regime we use the \textsc{pySB} default options of \citet{Vink2021,Bjorklund2021,Hawcroft2024}. For WRs and RSGs we add the prescriptions of \citet{SV2020,Sander2023} and \citet{Beasor2020,Yang2023} respectively. \Vcl\ is relatively insensitive to most changes among the considered OB and WR prescriptions, and we show these effects in Appendix \ref{sec:OB_WR}. In Fig. \ref{fig:mdots_rsg} we show the effects of using the \mdot\ prescriptions of \citet{Beasor2020,Yang2023} on \Vcl\ for $Z_\text{MW}$. 

\begin{figure}
    \centering
    \includegraphics[width=1\linewidth]{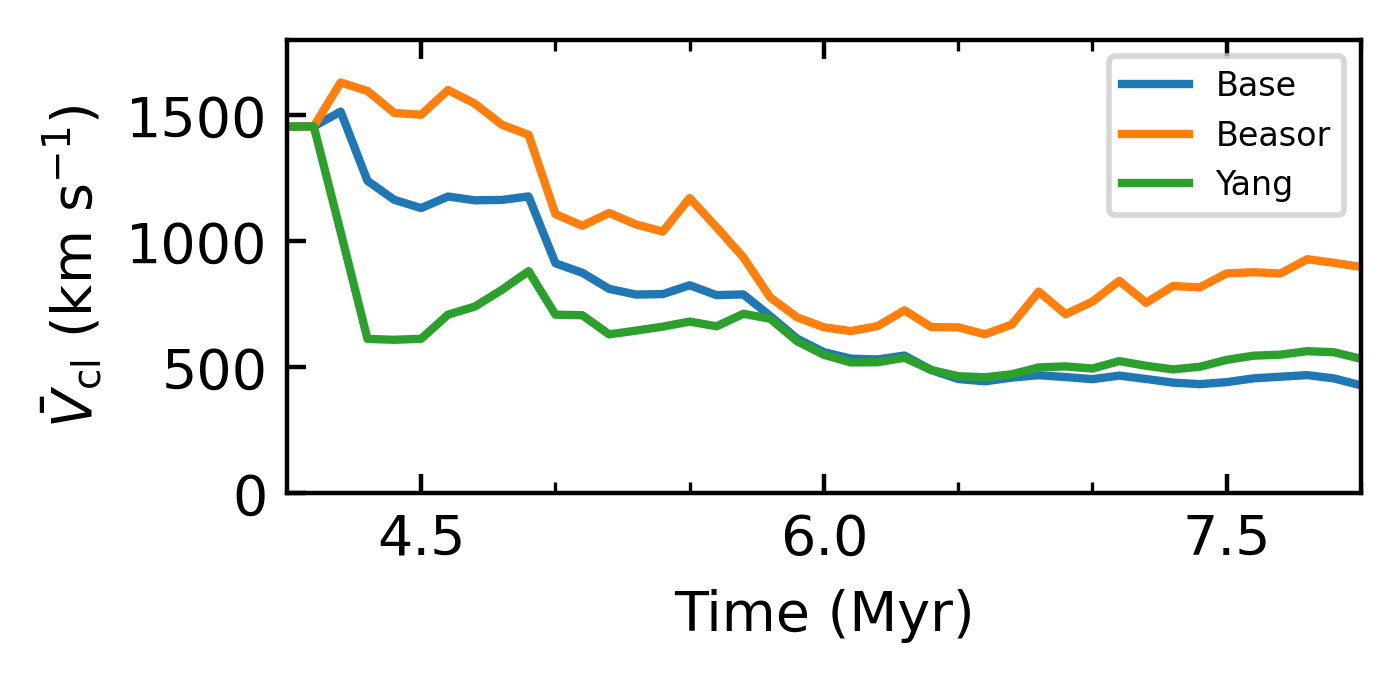}
    \caption{Time evolution of \Vcl\ (lower panel) for different RSG \mdot\ prescriptions at $Z_\text{MW}$.}
    \label{fig:mdots_rsg}
\end{figure}

Between $\sim4-5.5$\,Myr the difference in $V_{\mathrm{cl}}$ from changes in assumptions of RSG \mdot\ is of order $500-1000$\kms. These prescriptions are independent of $Z$, and in general RSG \mdot\ does not depend much on $Z$ \citep[e.g.,][]{Beasor2020,Antoniadis+2025}. Thus CSG mass-loading is likely to be consequential for YMCs in general.

\section{Discussion \& Conclusions}
\label{sec:discussion}

In this work, we present a first estimate of the effects of stellar evolution on the average velocity of a YMC wind. There are some effects we have not considered that are expected to play a role. We use the formalism of \citet{Canto2000}, which assumes the stellar winds interact with maximum efficiency, producing a uniform cluster wind, and we do not account for IMF stochastic effects either. Some parts of the cluster wind can move faster, and the velocity profile is expected to deviate from spherical symmetry \citep{Haerer2023,Vieu2024}. YMCs are embedded in their parent molecular cloud until around 1 Myr and also expel the left-over natal gas in this timeframe \citep{PortegiesZwart2010}, and we do not account for this here. This leftover material will mass-load the cluster wind at early times, reducing \Vcl.

    We also assume only single star evolution. Most massive stars will experience some binary interaction over their lifetime \citep[e.g.,][]{Sana2012}, and this can lead to effects that would both increase \Lmecht\ (binary stripping creating more WR-like stars) and increase mass-loading (e.g., ejection of common envelopes and luminous red novae). The net result of these competing effects is unclear. We adopt $M_{\mathrm{init}}$~= 120\Msun. Very massive stars (VMSs) above this limit are likely important for 1-2~Myr, but will have died off by the time where winds of WRs and CSGs become important \citep{Sabhahit2022}. We also do not consider the effects of SNe. It is not clear whether stars above $\sim30$\Msun\ explode, and what dependence this maximum SN mass may have on $Z$ and \mdot\ prescription. We only consider single stellar populations (SSPs). Given the expected sensitivity to number, initial masses, and ages of sub-populations, we defer this to future work, where we will also include binarity, VMSs and SNe.

Nevertheless, we derive important insights from our study. Clearly, assuming only WR stars contribute to \Mdotcl\ can overestimate \Vcl\, and that material expelled in cooler stages will be present in significant quantities for all times that WR stars are active. In a sufficiently populated cluster, \Mdotcl\ will never be dominated only by WR stars (unlike \Lmecht), so our general trend should hold even when assuming different stellar wind prescriptions. \citet{Morlino2021} assume \Vcl\ of 3000\kms\ to explain acceleration of PeV energy cosmic rays by YMC cluster winds, which is likely an overestimate for Galactic YMCs older than $\sim2$Myr. CCSNe could instead accelerate up to these energies \citep[e.g.,][]{Vieu2023,Haerer2025b}.

The qualitative difference in \Vcl\ between $Z_\text{LMC}$ and $Z_\text{SMC}$ is important for interpreting feedback from observations. At high $Z$, when considering all contributors, our findings are in broad agreement with currently used values \citep[e.g.,][]{Kader2025}. However, for $Z_\text{SMC}$ and below it is clear that \Vcl$\sim$500-1000\kms is more appropriate in the pre-SN feedback regime before $\sim$5-10\,Myr \citep{Schneider2021}. The sensitivity of our results to CSG \mdot\ assumptions underlines the need for empirical measurements not affected by dust-to-gas ratio assumptions in a range of environments \citep{Beasor2020,Decin2024}, e.g., via ram-pressure balancing as proposed by \citet{Larkin2025}. CSG mass loading likely impedes YMC outflows across all $Z$. Yet, current galaxy evolution simulations will struggle to resolve contributions from individual stars as the typical cell resolutions are a few $M_\odot$ \citep[e.g.,][]{Lahen2025}. We conclude:
\begin{itemize}
    \item The average cluster wind velocity \Vcl\ shows a strong $Z$ dependence, with the biggest difference obtained between $Z_\text{LMC}$ and $Z_\text{SMC}$. We advise assuming \Vcl$\sim$500-1000\kms\ for clusters at $Z_\text{SMC}$ and lower.
    \item By the time SNe start dominating the energetics of stellar feedback from clusters ($\sim$5-10\,Myr), the stellar wind $L_{\mathrm{mech}}$ cannot drive a cluster wind with $\Vcl>500$\kms, at least for the SSPs studied here and for $Z \geq Z_\text{LMC}$.
    \item Our findings support the canonical assumption that \Lmecht\ is well approximated by assuming only the WR stars of a cluster contribute, but we find \mdot\ from CSGs mass-load the cluster wind and can slow \Vcl\ by $500-2000$\kms.
    \item \Vcl\ is particularly sensitive to current uncertainties in CSG \mdot\ prescriptions. We find that changes in RSG prescriptions produce differences of order 1000\kms\ in \Vcl, highlighting the impact of uncertainties in RSG \mdot\ rates on understanding feedback from YMCs for all $Z$.
\end{itemize}

\begin{acknowledgements}
 CJKL gratefully acknowledges support from the International Max Planck Research School for Astronomy and Cosmic Physics at the University of Heidelberg in the form of an IMPRS PhD fellowship.
RRL, AACS, and CJKL are supported by the Deutsche Forschungsgemeinschaft (DFG, German Research Foundation) in the form of an Emmy Noether Research Group – Project-ID 445674056 (SA4064/1-1, PI Sander). This project was co-funded by the European Union (Project 101183150 - OCEANS).
This research has made use of the Astrophysics Data System, funded by NASA under Cooperative Agreement 80NSSC21M00561. This study used these software packages: 
Numpy \citep{HarMilVan20}, matplotlib \citep{Hun07}.
The authors thank B. Reville, C. Leitherer and N. Lah\'en for comments on the manuscript and the anonymous referee for their constructive feedback which has improved the manuscript. We also thank the organisers of the Topical Overview on Star Cluster Astrophysics workshop, where parts of this work were conceived.\\
\textit{Contributions. CJKL led the project and modified \textsc{pySB} to include the new stellar wind prescriptions. CH developed \textsc{pySB} and assisted with modifications. CJKL and JM conceived the project. RRL, LH and AACS provided theoretical support for the project. All authors contributed to drafting the manuscript.}    
\end{acknowledgements}

\bibliographystyle{aa} 
\bibliography{biblio.bib} 

\appendix

\section{Sensitivity to OB and WR wind prescriptions}\label{sec:OB_WR}

\subsection{OB winds}
\begin{figure}[!htbp]
    \centering
    \includegraphics[width=1\linewidth]{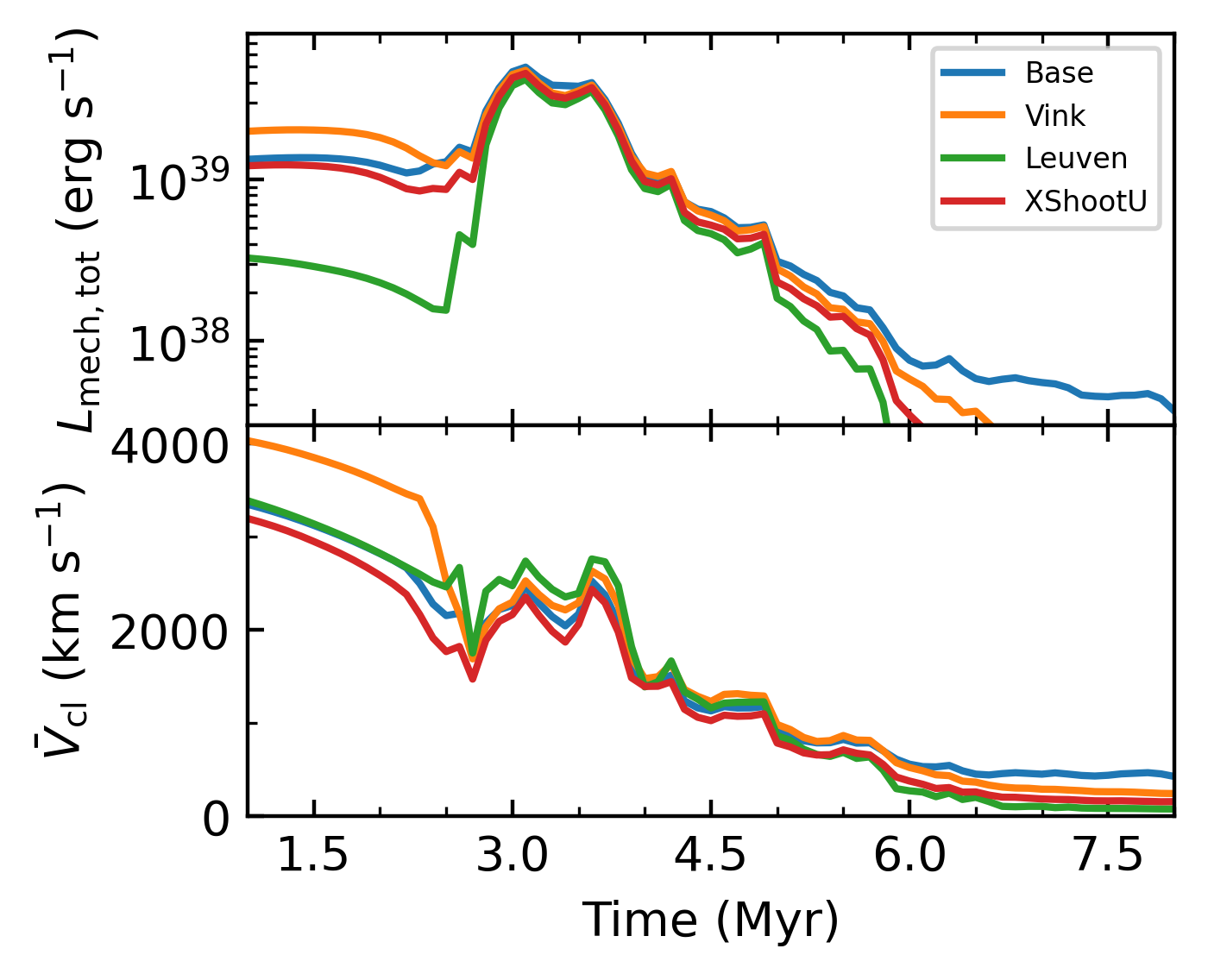}
    \caption{Time evolution of \Lmecht\ (upper panel) and \Vcl\ (lower panel) for different OB wind parameter prescriptions at $Z_\text{MW}$.}
    \label{fig:mdots_OB}
\end{figure}

In Fig. \ref{fig:mdots_OB} we show the effects of using the wind prescriptions of \citet{Vink2021,Bjorklund2021,Hawcroft2024} on \Lmecht\ and \Vcl\ for $Z_{MW}$.

For \Lmecht\, the prescriptions introduce variations of 0.1\,dex or less, with the exception of the Bj{\"o}rklund rates which reduce \Lmecht\ by $\sim$ 0.8 dex, as discussed in \citet{Hawcroft2025}. The variations are only prominent before evolved stars begin to contribute at $\sim2.5$\,Myr.

For \Vcl\, variations until this time are present but of order 100 \kms\, with the exception of the Vink rates. The increased \vinf\ in this prescription increases \Vcl\ by $\sim500$ \kms\ before evolved stars begin to contribute. In both cases, the changes in OB wind prescriptions are negligible after this time.

\subsection{WR winds}
\begin{figure}
    \centering
    \includegraphics[width=1\linewidth]{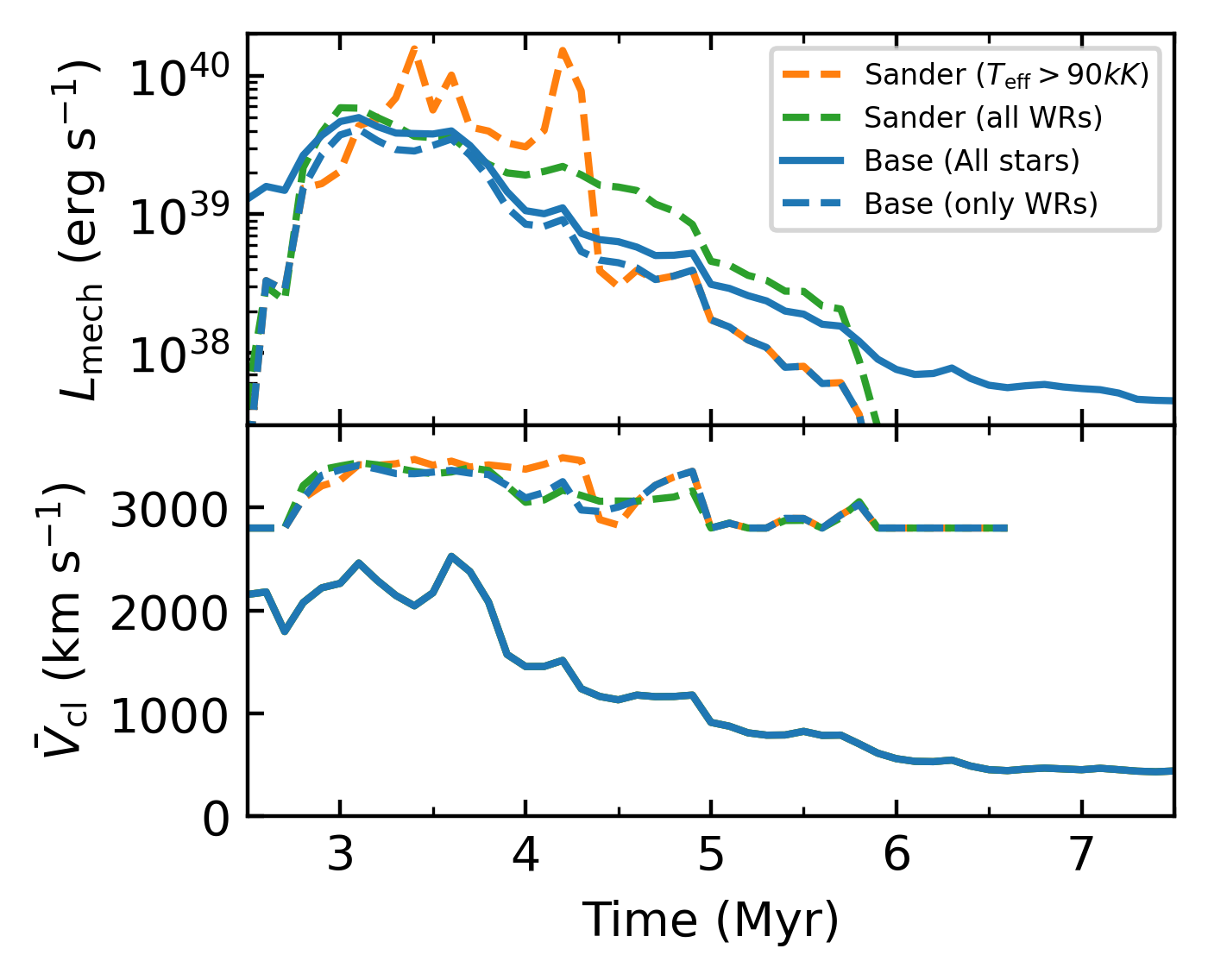}
    \caption{Time evolution of $L_{\mathrm{mech}}$ (upper panel) and \Vcl\ (lower panel) for $Z_\text{MW}$, assuming the base prescription for all stars (blue solid), base prescription for only WR stars (blue dashed), the Sander prescription with \teff\ dependence for WR stars with \teff$>$90kK (orange dashed), and the Sander prescription without \teff\ dependence for all WR stars (green dashed)}
    \label{fig:all_vs_WR_sander}
\end{figure}

In Fig. \ref{fig:all_vs_WR_sander} we show the effects of using the WR wind prescription of \citet{SV2020}, including the temperature dependence recommended in \citet{Sander2023}, on \Lmecht\ and \Vcl\ for $Z_\text{MW}$, assuming only WR stars contribute as for Fig. \ref{fig:all_vs_WR}.  

We test this prescription for two cases, firstly applying only the \citet{SV2020} prescription to all WR stars as defined earlier, and secondly applying the \citet{SV2020} prescription with temperature dependence \citep{Sander2023} to WR stars with \teff$>$90kK, which is the regime they were derived for, and using the Base prescription for cooler WR stars. We use the default \vinf\ prescriptions in \textsc{pySB} for all cases. 

In the upper panel, there is a moderate change in $L_{\mathrm{mech}}$ of order 0.1-0.2\,dex until $\sim$3\,Myr. At this point the hot WR stars become active and $L_{\mathrm{mech}}$ increases by up to an order of magnitude for the temperature-dependent case compared to the Base scenario until $\sim$$4.5$\,Myr. The effects of the WR prescriptions on \Vcl\ are comparatively small, changing by $\sim$$100$\kms\ at most.

There is a tiny contribution from winds classified as WR in the code for $Z_\text{IZw18}$ due to a ``sweet spot'' of low $Z$ and high \teff. We assess the impact of this by comparing the Base prescription with increasing the WR \teff\ criterion to $\log_{10} T_{\mathrm{eff}} > 4.44$ (i.e., from $\sim$25\,kK to $27.5$\,kK), thus making the WR assignment negligible. This makes a difference of order a few \% for \Lmecht\ and less than 1\% for \Vcl.

\end{document}